\documentstyle[11pt,paspconf]{article}

\begin{document}

\title{Future millimeter, submillimeter and far-infrared surveys and their 
successful follow up}

\author{A. W. Blain}
\affil{Institute of Astronomy, Madingley Road, Cambridge, CB3 0HA, UK} 

% Notice that some of these authors have alternate affiliations, which
% are identified by the \altaffilmark after each name.  The actual alternate
% affiliation information is typeset in footnotes at the bottom of the
% first page, and the text itself is specified in \altaffiltext commands.
% There is a separate \altaffiltext for each alternate affiliation
% indicated above.

% The abstract is entered in a LaTeX "environment", designated with paired
% \begin{abstract} -- \end{abstract} commands.  Other environments are
% identified by the name in the curly braces.

% Poster authors ONLY may omit the abstract in order to gain a little
% more page space for the text of the poster.

\begin{abstract}
The first unbiased surveys for high-redshift galaxies in the submillimeter
(submm) and far-infrared (FIR) wavebands have been made over the last two 
years. When combined with the intensity of extragalactic background radiation 
in the same wavebands, we have a reasonable first view of the dust 
enshrouded Universe. Limited spatial resolution hampers the identification of 
the submm galaxies, and follow-up observations currently consume much 
larger amounts of telescope time than the initial submm-wave detections. 
Here the prospects for future submm/FIR surveys are discussed, highlighting 
the importance of planning careful multiwaveband radio/submm/FIR 
observations. These can be used to yield photometric redshift information. 
Radio observations are vital to give accurate positions for the detected sources. 
\end{abstract}

% Keywords should be included, but they are not printed in the hardcopy.


% That's it for the front matter.  On to the main body of the paper.
% We'll only put in tutorial remarks at the beginning of each section
% so you can see entire sections together.

\section{Introduction}

The history of 
dust obscured energy production in galaxies, whether due either to 
star-formation activity or to accretion onto active galactic nuclei (AGN), 
has been 
investigated (Guiderdoni et al.\ 1998; Blain et~al.\ 1999a,c; Tan et al.\ 1999; 
Trentham et al.\ 1999), using data from a 
combination of surveys for high-redshift galaxies at 850\,$\mu$m using 
the SCUBA camera (Holland et al. 1999) (Smail et al.\ 1997; 
Barger et al.\ 1998; Hughes et al.\ 1998; Barger et al.\ 1999a; Blain et al.\ 1999b; 
Eales et al.\ 1999) and the determination of the submm/far-infrared (FIR) 
extragalactic background radiation intensity (Puget et al.\ 1996; Fixsen et al.\ 
1998; Hauser et al.\ 1998; Schlegel et al.\ 1998). 
The bulk of the mm/submm-wave background radiation 
intensity at wavelengths longer than 500\,$\mu$m has now been resolved into 
discrete sources 
(Blain et al.\ 1999c). Existing measurements of the counts and background 
radiation intensity contributed by distant dusty galaxies are shown in 
Figs 1 \& 2. In Fig.\,1 counts derived using a well-constrained model 
(Blain et al.\ 1999b) are also shown. The confusion noise expected in
surveys from dusty galaxies (Blain et al.\ 1998a,b), radio galaxies 
(Toffolatti et al.\ 1998) and the ISM (Helou \& Beichman 1990) are shown in Fig.\,3. 

However, less than 100 submm-selected galaxies are currently known, 
and redshifts and high-quality 
multiwaveband data available for very few.  
%(Hughes et al.\ 1998; Smail et al.\ 1998; Barger et al.\ 1999b; Lilly et al.\ 1999). 
The confirmation of their optical counterparts is currently a very time consuming 
process. First, an extremely deep VLA radio image is desirable, to give a 
sub-arcsec position for the submm galaxy. An optical redshift must then be 
found (e.g. Ivison et al.\ 1998), before a mm-wave interferometric 
detection of CO molecular line emission at the same redshift confirms 
the identification unequivocally (Frayer et al.\ 1998, 1999). 
The observing time 
involved in this identification process is up to ten times greater than that 
required to detect each submm-luminous galaxy. In future 
mm/submm-wave surveys, using for example the wide-field BOLOCAM 
bolometer array camera (Glenn et al.\ 1998), the detection rate of galaxies should
increase from a few per night to several per hour. Hence, even 
with access to the very finest facilities, this follow-up procedure will be 
completely inadequate to keep up with the growing catalog of submm galaxies. 

Here we discuss how observations of dusty galaxies using future facilities over 
the next few years -- in particular BOLOCAM, 
the long-wavelength MIPS instrument on {\it SIRTF} and the upgraded VLA -- 
can be combined to provide redshift information without recourse to optical 
telescopes. Despite the large detection rate, the classification and 
identification of the most interesting detected galaxies for detailed follow-up 
observations should be able to keep pace. 

\begin{figure}[t]
\begin{center}
\plotfiddle{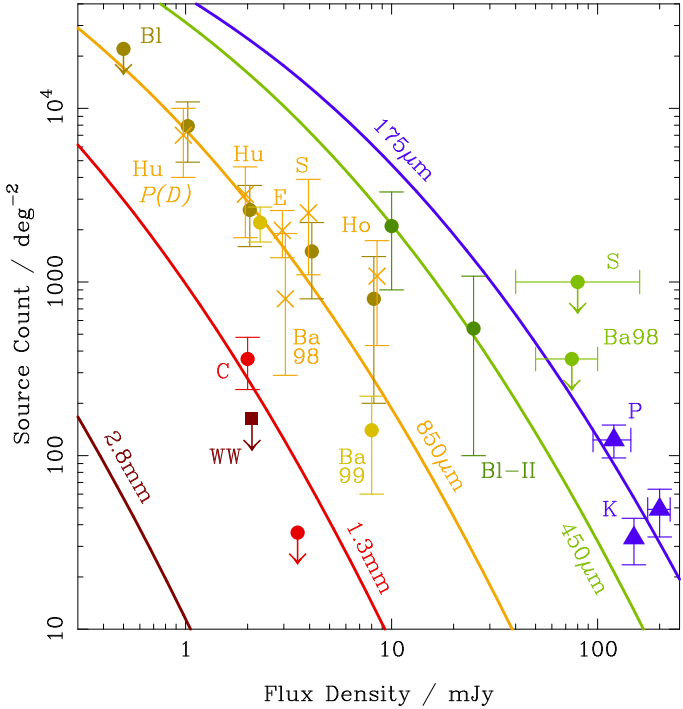}{8.1cm}{-90}{60}{60}{-225}{315}
\end{center}
\caption{Counts of mm, submm and FIR galaxies. Data: Blain et~al.\ (1999b, 
2000) 
-- Bl, Bl-II; 
Hughes et al.\ (1998) -- Hu; Eales et al.\ (1999) - E; Smail et~al.\ (1997) 
-- S; Barger et al.\ (1999a) -- Ba99; Barger et~al.\ (1998) -- Ba98; Holland 
et al.\ (1998) -- Ho; Carilli et al. (2000) -- C; 
Wilner \& Wright (1997) -- WW; 
Puget et~al.\ (1999) -- P; and Kawara et al.\ (1998) -- K. The results at 
2.8\,mm (WW) are from BIMA, at 1.3\,mm from IRAM, 
at 850 and 450\,$\mu$m from SCUBA and at 
175\,$\mu$m 
from {\it ISO}. The lines are taken from the `modified gaussian model', 
a modified version of the Blain et al.\ (1999b) `Gaussian model', which accounts 
for the preliminary redshift distribution of SCUBA galaxies (Barger et al.\ 1999b).   
} 
\end{figure} 

\begin{figure}[t]
\begin{center}
\plotfiddle{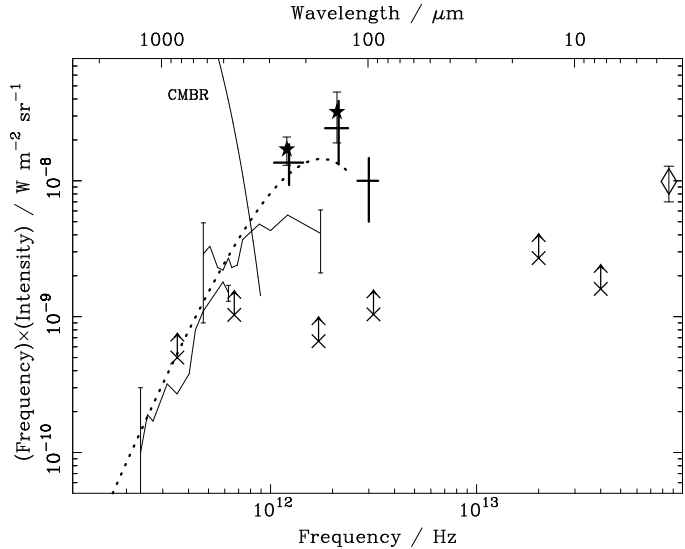}{5.9cm}{-90}{40}{40}{-150}{215}
\end{center}
\caption{The background radiation intensity 
from the mm to near-IR wavebands. Solid line -- Puget et al.\ 
(1996). Dotted line -- Fixsen et~al.\ (1998). Diagonal crosses  
(from left to right) -- Blain et al.\ (1999b, 2000); Ivison et al.\ (1998, 
2000); 
Kawara et al.\ (1998) and Puget et~al.\ (1999); and (last pair) Altieri et al.\ (1999). 
Stars -- Schlegel et al.\ (1998). Vertical crosses -- Hauser et al.\ (1998) 
and Dwek et al.\ (1998). Diamond -- Dwek \& Arendt (1998).}
\end{figure}     

\section{The properties of luminous high-redshift dusty galaxies} 

Three galaxies detected in submm surveys have sure identifications, as well 
as two cD galaxies detected in clusters (Edge et al.\ 1999). These are 
ultraluminous galaxies at redshifts $z=1.06, 2.56$ and 2.81 (Soucail et al.\ 1999; 
Frayer et al.\ 1998, 1999; Ivison et al.\ 1998, 2000). An important goal is to 
extend the size of the sample of submm-selected 
galaxies, and to identify a larger number. A wide range of powerful mm, submm 
and FIR instruments that are suitable for carrying out surveys 
(see Table\,1) will become available over the next decade. Some of their 
specifications are listed in Table\,1, along with the rate at which these 
instruments should be able to survey the sky and the depth at which they 
are expected to be affected by confusion noise. In Fig.\,4 the detection rate 
of galaxies that could be achieved using these instruments are shown. 

Note that while the Atacama Large Millimeter Array (ALMA) is a very powerful 
survey instrument, it will be most important for making resolved images of 
the galaxies detected using alll the other facilities. The {\it Planck Surveyor} 
microwave background imaging satellite will also provide a very valuable 
all-sky survey at wavelengths between 10\,mm and 350\,$\mu$m. 

Detection rates of order 100\,hr$^{-1}$ are expected in future 
mm/submm-wave surveys, probing galaxies at $z > 1$, as compared with the 
current rate of about 0.2\,hr$^{-1}$. Is it possible to make 
multiwaveband mm/submm/FIR 
observations using a combination of the instruments listed in 
Table\,1, and to generate photometric redshifts for the detected sources? 

\begin{figure}[t]
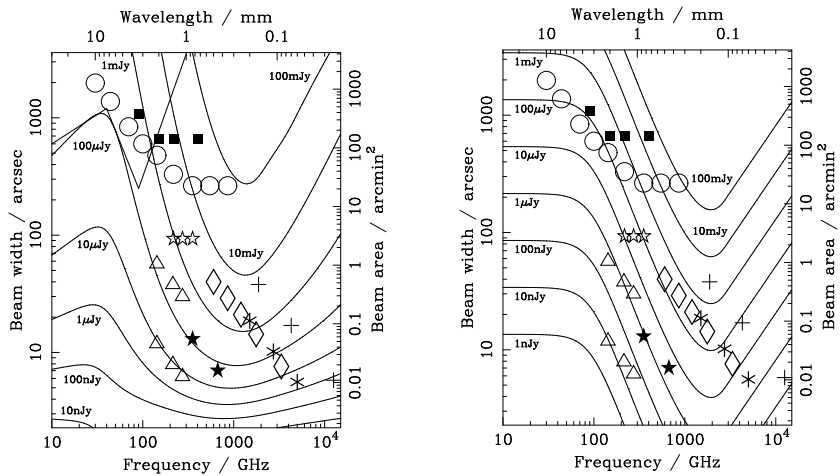
 
\begin{center}
\plotfiddle{blainatalk3a.ps}{5.1cm}{-90}{36}{36}{-230}{180}
\end{center}
\begin{center} 
\plotfiddle{blainatalk3b.ps}{4.3cm}{-90}{36}{36}{-60}{328}
\end{center} 
\vskip -5cm 
\caption{1-$\sigma$ confusion noise expected in mm to FIR surveys. Left: 
due to dusty external galaxies (extended from Blain et al.\ 1998a,b).  
Radio-loud objects may make a significant contribution to the top left of the 
jagged solid line (Toffolatti et al.\ 1998). Right: due to the Milky Way ISM, if 
the surface brightness $B_0= 1$\,MJy\,sr$^{-1}$ at 100\,$\mu$m: see Helou \& 
Beichman (1990), and data from Gautier et al.\ (1992), D\'esert et al.\ (1990) 
and Finkbeiner et al.\ (1999). The ISM confusion noise scales as $B_0^{1.5}$. 
The extrapolations to arcsecond scales are potentially uncertain. 
Experiments are shown by: circles -- {\it Planck Surveyor}; squares -- 
BOOMERANG; empty stars -- SuZIE; triangles -- BOLOCAM (CSO above; 
LMT below); 
filled stars -- SCUBA; diamonds -- {\it FIRST}; asterixes -- SOFIA; 
crosses -- {\it SIRTF}. ALMA and {\it SPECS} have resolution limits below 
the bottom of the frames. 
}
\end{figure}

\begin{table}[t] 
\caption{
Wavelengths $\lambda$, sensitivities (as noise equivalent 
flux density -- NEFD), fields of view (FOV), mapping speeds and confusion 
limits due to galaxies and the ISM for future surveys. The speed presented is 
the mapping rate required to reach a 5-$\sigma$ detection at a depth of 10\,mJy, 
which depends on the square of the detection flux density limit. The confusion 
noise is defined as the flux at which there is one source per beam (see Fig.\,3). 
The ISM confusion (in brackets) is calculated for a 100-$\mu$m surface 
brightness $B_0$=1\,MJy\,sr$^{-1}$ and scales as $B_0^{1.5}$.
} 
\begin{center}\scriptsize
\begin{tabular}{lccccc}
Name & $\lambda$ & NEFD & FOV & Speed & Confusion\\ 
& ($\mu$m) & (mJy/$\sqrt{Hz}$) & (arcmin$^2$) & (deg$^2$\,hr$^{-1}$) & (mJy)\\ 
\tableline
SCUBA+ & 850 & 40 & 1.7 & $4.3 \times 10^{-3}$ & 0.12 ($9 \times 10^{-4}$) \\  
       & 450 & 120 & 1.7 & $4.8 \times 10^{-4}$ & 0.053 ($3 \times 10^{-3}$) \\ 
SCUBA-II{\tablenotemark{i}}& 850 & 50 & 64 & 2.6 & 0.12 ($9 \times 10^{-4}$) \\
       & 450 & 200 & 64 & 0.16 & 0.053 ($3 \times 10^{-3}$) \\
SOFIA{\tablenotemark{a}} & 200 & 408 & 9.0 & $2.8 \times 10^{-4}$ & 1.2 (0.30) \\
BOLOCAM-CSO{\tablenotemark{b}} & 
1100 & 42 & 44 & 0.10 & 0.32 ($2 \times 10^{-3}$) \\
BOLOCAM-LMT{\tablenotemark{b}} & 
1100 & 2.8 & 2.5 & 1.6 & $6 \times 10^{-3}$ 
($4 \times 10^{-5}$)\\ 
{\it FIRST}-SPIRE{\tablenotemark{c}} & 
500 & 114 & 40 & $1.2 \times 10^{-2}$ & 2.9 (0.16) \\
	          & 350 & 90 & 40 & $2.0 \times 10^{-2}$ & 2.6 (0.12) \\
	          & 250 & 84 & 40 & $2.3 \times 10^{-2}$ & 1.6 (0.24) \\
{\it FIRST}-PACS{\tablenotemark{c}}  & 
170 & 1.6 & 16 & $6.9 \times 10^{-3}$ & 0.45 (0.16) \\
	          & 90 & 3.6 & 16 & $1.4 \times 10^{-3}$ & 0.01 (0.01) \\ 
{\it SIRTF}-MIPS{\tablenotemark{d}}  & 
160 & 18 & 2.5 & $3.1 \times 10^{-2}$ & 6.6 (3.1) \\
	          & 70 & 4.5 & 25 & 5.0 & 0.28 (0.07) \\
		  & 24 & 1.8 & 25 & 31 & $6 \times 10^{-4}$ 
($2 \times 10^{-4}$) \\
ALMA{\tablenotemark{e}} 
& 870 & 1.1 & 0.07 & 0.24 & $< 10^{-7}$ ($< 10^{-6}$) \\
 & 450 & 15 & 0.02 & $3.7 \times 10^{-4}$ & $< 10^{-8}$ ($< 10^{-4}$) \\
{\it SPECS} testbed{\tablenotemark{f}} & 
250 & 0.17 & 4 & 0.46 & $\sim 10^{-5}$ ($\sim 10^{-3}$) \\
{\it SIRTF}-IRAC{\tablenotemark{d}} & 
8.0 & 0.15 & 25 & 0.44 (0.1\,mJy) & $< 10^{-6}$ ($\sim 10^{-6}$) \\ 
& 5.8 & 0.10 & 25 & Simultaneous & Similar to above \\
& 4.5 & 0.022 & 25 & Simultaneous & Similar to above \\ 
& 3.6 & 0.022 & 25 & Simultaneous & Similar to above \\ 
Upgraded VLA & 20.5\,cm & 0.40 & 700 & 2.0 (0.1\,mJy) & ... (...) \\
\end{tabular}
\vskip -0.5cm 
\end{center}
\tablenotetext{a}{Davidson et al.\ (1999); http://sofia.arc.nasa.gov/}
\tablenotetext{b}{Glenn et al.\ (1998); 
http://www-lmt.phast.umass.edu/ins/continuum/bolocam.html} 
\tablenotetext{c}{Bock (1999)}
\tablenotetext{d}{http://ssc.ipac.caltech.edu/sirtf/}
\tablenotetext{e}{http://www.alma.nrao.edu} 
\tablenotetext{f}{Mather et al.\ (1998); http://www.gsfc.nasa.gov/astro/specs/}
\tablenotetext{i}{http://www.jach.hawaii.edu/JACpublic/JCMT/scuba/scuba2} 
\end{table}

\begin{figure}[th]
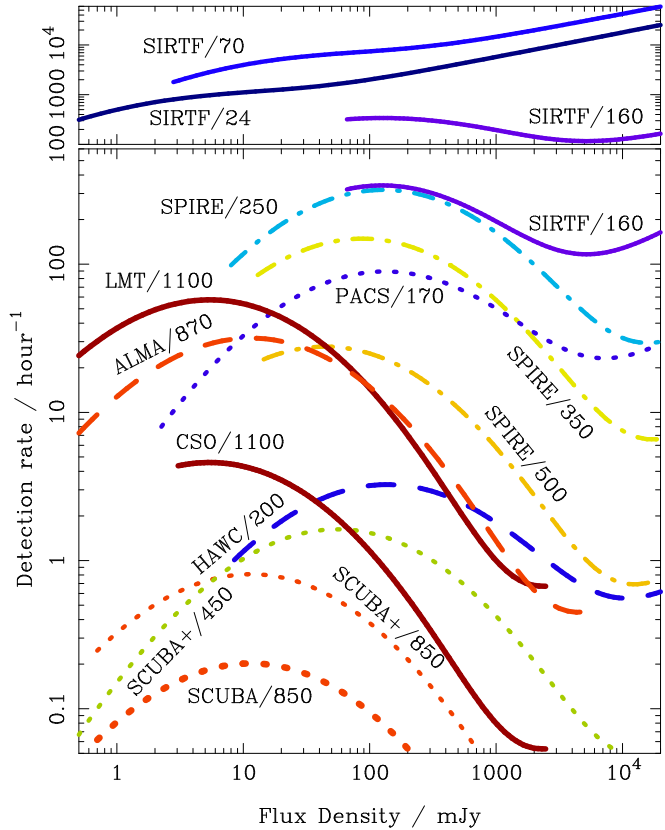

\begin{center}
\plotfiddle{blainatalk4a.ps}{1.1cm}{-90}{58}{58}{-225}{204}
\end{center}
\begin{center}
\plotfiddle{blainatalk4b.ps}{7.5cm}{-90}{58}{58}{-225}{300}
\end{center}
\caption{The detection rate as a function of 5-$\sigma$ depth in a range of 
surveys. Numbers are the observing wavelengths in $\mu$m. HAWC is the 
FIR camera for SOFIA; SPIRE (PACS) is the long-(short-)wave 
imaging instrument for {\it FIRST}; BOLOCAM will operate at 
the CSO and the 50-m LMT. The curves stop at about the confusion 
limit (left) and where the count falls below $1/4\pi$\,sr$^{-1}$ (right).} 
\end{figure} 

\section{Prospects for submm/FIR photometric redshifts} 

A common broad-band spectral energy distribution (SED) can describe 
most submm-selected galaxies: see Fig.\,1 in Blain (1999; this volume).
There are several possible ways to derive photometric redshifts from radio, mm, 
submm and FIR photometry assuming this SED, which has a very generic form 
at wavelengths longer than about 60\,$\mu$m. 

\begin{enumerate} 
\item The radio--submm spectral index. As discussed by Carilli \& Yun (1999; 
2000) 
and Blain (1999; this volume), the spectral break between non-thermal 
synchrotron emission 
and thermal dust emission, which typically occurs at a restframe wavelength of 
about 3\,mm, can be used to provide redshift information. The quantity $(1+z)/T$, 
where $T$ is the dust temperature can be determined reasonably accurately 
using this method: see Smail et al. (1999) for an application. 
\item The position of the peak of the dust SED. Measurements at wavelengths 
on both sides of the redshifted peak of the dust SED can be used to fix 
the quantity $(1+z)/T$. This requires joint submm and FIR observations. See 
also Hines (1999; this volume). 
\item The submm--optical flux density ratio. Distant dusty galaxies seem to 
show a wide range of optical properties (Smail et al.\ 1998; Barger et al.\ 1999b; 
Ivison et al.\ 2000), but more distant galaxies should typically be optically 
fainter, because of the very different $K$-corrections in each band. This
information could be used to provide a coarse indication of redshift. 
\item Mid-IR (MIR) and near-IR spectral features can be exploited to 
provide redshift information. This is outside the scope of this paper: see 
Xu et al.\ (1998) and Simpson \& Eisenhardt (1999) for more information.  
\end{enumerate}

The issue of cross identification between surveys at different wavelengths 
should not be too challenging. At the relevant wavelengths of 1.1\,mm 
(CSO-BOLOCAM) and 160, 70 and 24\,$\mu$m ({\it SIRTF}-MIPS), the angular 
resolution of the survey maps will be well matched, at about 30, 45, 20 and 
7\,arcsec respectively. The sub-arcsec positional accuracy of a VLA detection 
will allow a counterpart source to be sought in deep optical and near-IR 
images. 

\section{The best prospects for future surveys} 

\begin{figure}[t]
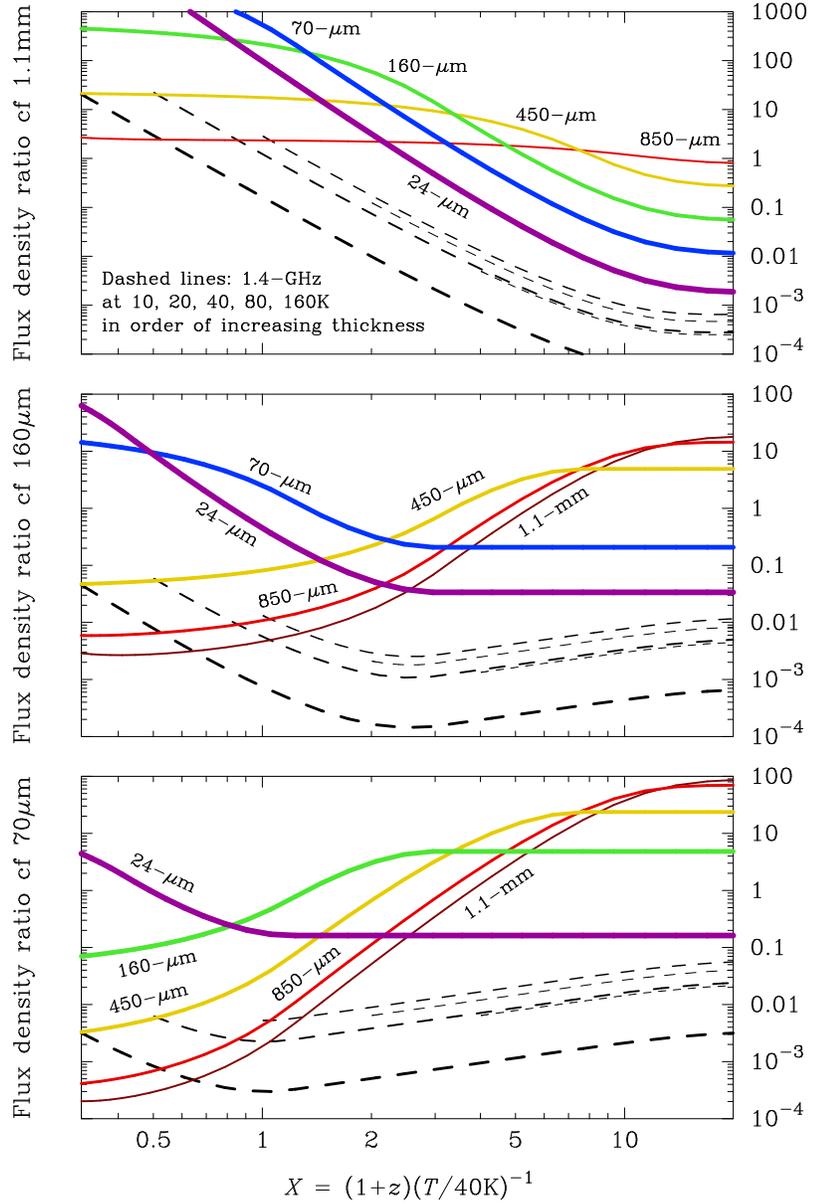

\begin{center}
\plotfiddle{blainatalk5a.ps}{4.2cm}{-90}{50}{50}{-195}{230}
\end{center}
\begin{center}
\plotfiddle{blainatalk5b.ps}{4.2cm}{-90}{50}{50}{-195}{230}
\end{center}
\begin{center}
\plotfiddle{blainatalk5c.ps}{4.2cm}{-90}{50}{50}{-195}{230}
\end{center}
\caption{Flux density ratios expected at a range of wavelengths 
in the radio, mm, submm and FIR wavebands, as compared with 
survey wavelengths of 1.1\,mm (top), 160\,$\mu$m (middle) and 
70\,$\mu$m (bottom), appropriate to BOLOCAM and {\it SIRTF}-MIPS surveys. 
If the curves have a slope, rather than being horizontal, 
then the ratio is a potentially useful photometric redshift indicator, if the 
source is sufficiently bright to be detected in both bands. Note that
little information is provided by a measurement in the 450/850-$\mu$m 
atmospheric windows unless $X > 5$. Hence, air-/space-borne 
observations at shorter wavelengths are required to constrain $X$ over the 
likely range of redshifts for sources detected in a mm-wave survey.
}
\end{figure} 

New instruments that will be coming into service over the next few years are 
listed in Table\,1. Of these the most timely and powerful are BOLOCAM, 
{\it SIRTF} and the upgraded VLA, which will cover the MIR/FIR, mm and 
radio wavebands. However, the disparate mapping speeds of each instrument 
mean that different combinations of overlapping surveys will be required. 
To best exploit the capabilities of {\it SIRTF} 
and BOLOCAM, there seem to be three clear survey strategies. The
photometric redshift information that can be derived from each survey 
is shown in the 3 panels of Fig.\,5. 

\begin{enumerate} 
\item A 1.1-mm BOLOCAM survey at the 10-m CSO mapping at the most 
efficient depth, 5-$\sigma \simeq 10$\,mJy (Fig.\,4). About 4 
galaxies per hour should be detected, 
and ultimately about 50\,hr$^{-1}$ at the 50-m LMT. These 
mm-wave surveys will be a very efficient way to detect 
high-redshift galaxies; almost all of the detected sources should be at $z>1$,
about 50\% at $z>2$, and about 10\% at $z>5$ (Barger et al.\ 1999a; Blain et al.\ 
1999c). 
\item A 160-$\mu$m {\it SIRTF} survey to an efficient detection
depth of 100\,mJy (Fig.\,4), surveying about 3\,deg$^2$ per hour and detecting 
about 300 galaxies per hour. This survey exploits {\it SIRTF}'s longest 
wavelength band to boost the number of high-redshift galaxies detected; 80\% 
of the detected sources are expected to be at $z>1$, and about 10\% at $z>2$.
\item A 70-$\mu$m {\it SIRTF} survey to a confusion limited depth of 3\,mJy. 
This will yield a tremendously large number of sources, at a rate of 
about 6000\,hr$^{-1}$. A shallower wider survey in this 
band would lead to an even greater detection rate: see Fig.\, 4. 
About 60\% of the detected galaxies are expected to be at $z>0.5$, and about 
5\% at $z>2$. 
\end{enumerate} 

\subsection{Following up and finding photometric redshifts} 

By making coupled FIR/MIR {\it SIRTF} observations of CSO-BOLOCAM
survey fields, the form of the SED of the detected galaxies will be constrained. 
At 160 and 70\,$\mu$m, about 9 and 23\% of the time 
spent surveying at 1.1-mm is required to reach the confusion limits of 
70 and 6\,mJy respectively at 5-$\sigma$ significance. At 1.4\,GHz, 
24\,$\mu$m and 8\,$\mu$m, using the upgraded VLA, MIPS and IRAC 
respectively, 10\% of the 1.1-mm integration time will yield deep, unconfused 
maps to a 5-$\sigma$ depth of 70\,$\mu$Jy, 2\,mJy and 0.15\,mJy respectively. 
At the top of Fig.\,5 the efficiency of combining all this data to generate 
photometric redshifts can be seen. The radio--submm flux density ratio will 
typically provide useful information on the value of 
$X=(1+z)(T/40\,{\rm K})^{-1}$ if $X < 3$, while the {\it SIRTF}-MIPS data is 
useful to $X < 4$. Note that the sources detected will be several times brighter 
than the existing SCUBA galaxies. Deep ALMA high-resolution imaging of 
sources not detected in the initial VLA survey can be carried out in 
1-min integration. 

Observations with the upgraded VLA can cover the field of a 160-$\mu$m 
{\it SIRTF} survey to a 5$\sigma$ sensitivity of 120\,$\mu$Jy in 100\% of the 
time allocated to {\it SIRTF}. At this sensitivity, detections should be obtained
for all values of $X$ (see the middle of Fig.\,5); however, the shallow slope of 
the radio--FIR flux density ratio means that the photometric redshift 
information derived will be relatively noisy. From Fig.\,5, 5-$\sigma$ sensitivities 
of about 20 and 3\,mJy are required to detect a 100-mJy 160-$\mu$m source 
for interesting values of $X$. Maps to this sensitivity can be obtained in the 
70- and 24-$\mu$m MIPS channels in 100\% of the time allocated to the 
160-$\mu$m survey. mm-wave BOLOCAM observations cannot be made over 
the whole 160-$\mu$m {\it SIRTF} survey field to a depth sufficient to detect 
160-$\mu$m sources at interesting flux densities in a reasonable time, even 
at the LMT. 

Although the detection rate of galaxies in a 70-$\mu$m {\it SIRTF} survey 
will be very large, the survey will be impossible to follow-up in its entirety
at other mm/FIR wavelengths until the arrival of the ALMA and {\it SPECS}. 
However, in a time-matched survey, the VLA could reach a 5$\sigma$ depth of 
4.7\,$\mu$Jy, detecting and providing an accurate position for {\it SIRTF}
70-$\mu$m sources for all values of $X$ (see bottom of Fig.\,5). A 
LMT-BOLOCAM survey could cover the same area to a 5-$\sigma$ depth 
of 0.5\,mJy in the same integration time, sufficient to provide redshift 
information if $X > 2.5$.  

\subsection{Other instruments} 

The ALMA has excellent sensitivity, and so follow-up imaging 
observations to resolve the detected sources will be very rapid 
(see Blain 2000). This will 
also be  
true for the {\it SPECS} space interferometer. The {\it SIRTF}-IRAC 
camera, operating at its longest wavelength of 8\,$\mu$m will also be a very 
powerful instrument for detecting distant galaxies. However, because the 
models here are tuned to make predictions at longer wavelengths, refer to 
Eisenhardt (1999; this volume) 
for predictions of the performance of IRAC. 
%Ultra wide-band mm-wave 
%spectrometers may also allow spectroscopic redshifts to be found by observing 
%adjacent CO rotational emission lines in distant dusty galaxies.  

\section{Conclusions} 

The first generation of submm surveys have now resulted in significant 
understanding of dust obscuration in the high-redshift Universe, but only 
a few examples of well studied submm-selected star-forming galaxies and 
active galactic nuclei are known. Within the next few years BOLOCAM and 
{\it SIRTF} should dramatically increase the size of samples. By carefully 
combining the surveys made by these instruments and the upgraded VLA, 
photometric redshifts should be available for up to 90\% of the detected 
galaxies, thus providing information with which to test models of galaxy 
evolution and to prioritize follow-up observations. 

\acknowledgments

I thank Ian Smail, Rob Ivison and Jean-Paul Kneib for the success of the 
SCUBA lens survey and useful discussions. Also Lee Armus, Jaime Bock, 
Len Cowie, Jackie Davidson, Eli Dwek, Peter Eisenhardt, Jason Glenn, Dean 
Hines, Andrew Lange, Dave Leisawitz and OCIW.

% That's the end of the main body of the paper.  Now we will have some
% back matter.

% Now comes the reference list.  Since we typed out the citations ourselves,
% the reference list is enclosed in a "references" environment.  Each
% new reference begins with a \reference command which sets up the proper
% indentation.  Typography that may be required in the reference list by
% the editorial staff must be included by the author.
%
% Observe the "standard" order for bibliographic material: author name(s),
% publication year, journal name, volume, and page number for articles.
% Some journal names are available as macros; see the WGAS markup
% instructions for a listing of which ones have been "macro-ized".
% Note the use of curly braces to delimit the font changes: it is essential
% that this be done to limit the scope of the font declaration.
%
% There is no need to engage in any other typographic manipulation.

% That's all, folks.
%
% The technique of segregating major semantic components of the document
% within "environments" is a very good one, but you as an author have to
% come up with a way of making sure each \begin{whatzit} has a corresponding
% \end{whatzit}.  If you miss one, LaTeX will probably complain a great
% deal during the composition of the document.  Occasionally, you get away
% with it right up to the \end{document}, in which case, you will see
% "\begin{whatzit} ended by \end{document}".

\end{document}